# Observations of 't Hooft's sublattices and Dirac's monopole by inhomogeneous phases of solitons

Muhammad Imran Afzal[1], Hoonsoo Kang[1]* and Yong Tak Lee[1, 2]*

[1]Advanced Photonics Research Institute, Gwangju Institute of Science and Technology, Gwangju 500-712, Korea.
[2]School of Information and Communications, Gwangju Institute of Science and Technology, Gwangju, 500-712, Korea.
*hunskang@gist.ac.kr
*ytlee@gist.ac.kr

**Dirac theorized magnetic monopole[1]. Where magnetic flux leaked out as string singularity from north monopole. 't Hooft formulated generalized monopole in a compact space without the string singularity and predicted strongly bounded monopoles as torus subgroup or sublattice, from original gauge symmetry in nonAbelian grand unification theory[2] i.e., $G = SU(2) \rightarrow H = U(1)$. Where a defect monopole[3-4] may become a chromoelectric flux tube (dual superconductor). The tube can extend in straight line only[5-6]. However, in case of flux leakage (from the space) the 't Hooft's monopole is equivalent to Dirac's monopole[2]. Analogues of Dirac's monopoles and strings have been found in different mediums[7-11], however, even in synthetic gauge field[11] these are far from the original ideas[12]. The sublattice is also still elusive. According to Von Neumann and Wigner[13], bounding and anti-crossing occur between asymmetric and symmetric eigenstates, respectively. The former is experimentally demonstrated by resonance of photonic states and radiation losses[14]. Graphene-like resonance induces Anti-crossing[15], pseudomagnetic field and Landau levels in inhomogeneous strained waveguides[16]. While bound states are induced non-Hermicity (dissimilar radiation-loss rates)[17]. These experiments have paved the way for realizing optical analogues of weakly and strongly bounding of states, depending on strength of pseudomagnetic field. Here, we experimentally generated photonic graphene by resonance of inhomogeneously strained one dimensional lattices of triangular solitons. Where mildly twisted solitons are considered as north and south monopoles, while strongly twisted solitons considered as defect north monopoles. Weak bounding is observed between the opposite monopoles. Strong bounding occurred between the monopoles with same polarity, which is analogous to bounding in 't Hooft's torus sublattice**

**(dual superconductor), where a defect north monopole is transformed into the flux-like tube. Bogomolny's vortice-like symmetry[18-19] remained intact in all these observation. Dirac's north monopole along with the string is observed at one edge (end) of the lattice. The results presented in this paper were also described in terms of supersymmetry and quantum phase transitions including topological Anderson localization, and reported in ref [20].**

We considered a soliton with phase $\varphi_i$ as mildly twisted when $\varphi_{i+1} - \varphi_i > 0$ and $\varphi_i - \varphi_{i-1} > 0$, while $|\varphi_{i+1} - \varphi_i| \neq |\varphi_i - \varphi_{i-1}|$. Here we considered the mildly twisted solitons as monopoles. A soliton with phase $\varphi_i$ is treated as strongly twisted when $\varphi_{i+1} - \varphi_i < 0$ and $\varphi_i - \varphi_{i-1} > 0$ or $\varphi_{i+1} - \varphi_i > 0$ and $\varphi_i - \varphi_{i-1} < 0$. Former solitons are defect north monopoles and later are defect south monopoles. Owing to graphene-like resonance, mildly inhomogeneous strained-like modulation of phases $\varphi_i$ of the soliton lattices generated similar pseudomagnetic flux $\Phi$ owing to similar radiation rates, without undergoing the interband photonic transitions (evanescent tunnelling). The direction of the flux is opposite to y-axis. Due to similar radiation rate losses, the eigenstates are annihilated with similar rates in the same direction[20]. However, strong twists generated strong flux owing to exponential radiation rates. In this case, the direction of the flux or radiation is orthogonal (parallel to x-axis) to the flux $\Phi$ of mildly twisted solitons. These dynamics enabled us to observe Dirac's monopoles, Dirac's string at the end of the north monopole, 't Hooft's monopoles, the flux tubes and torus sublattices. In contrast with Dirac's string, the flux tubes are observed between monopoles of same polarities i.e. south monopoles. The south monopoles are connected with each other through color-like force via the flux tube. Originally color force is reminiscence of color charges i.e. quark and anti-quark. According to our best knowledge these observations have no experimental analogue, however theoretically predicted in a number of papers such as[1-2, 18-19,23] and present state of relevant progress is summarized in recent papers[3-4]. 't Hooft's type monopoles are also predicted by Polyakov[23]. Theory of Dirac's monopoles and 't Hooft's monopoles and torus sublattices have pivotal importance in magnetic and electrical charge quantization, grand unification theories, gravity and supersymmetry.

Our initial system consist of two lattices with strain-like phase modulation $\varphi_{ij}$, however both lattices have uniform lattice spacing, which is equal to 0.54 nm. Schematics of the lattices are depicted in Fig. 1. The solitons are imitated as north and south monopoles. Similar kind of symmetry has been considered in previous experiments of bound states with[16] and without self-magnetic effects[14, 21-22]. Green lines indicate monopoles which follow the symmetry of north, south monopoles. Orange lines symbolize the boundary monopoles. Red and purple lines indicate defect north and south monopoles (strongly twisted solitons), respectively. In Fig. 1 (a), there are five monopoles following the symmetry and two monopoles are defect north monopoles including two boundary monopoles. Second lattice of the monopoles is depicted in Fig. 1(b), where two of the monopoles are considered defect north monopoles, one defect south monopole and two boundary monopoles.

By graphene-like resonance of the lattices with themselves owing to passing through the experimental setup, Dirac's monopoles and the string are generated as schematically depicted in Fig. 2(a). Unmodulated magnetic flux is leaked out as string singularity from Dirac's north monopole only, which is represented as black curves in Fig. 2(a). On the other side, presence of a defect north monopole in monopoles symmetry i.e., a north monopole between two south monopoles, generated 't Hooft's predicted torus subgroup or sublattice. Here, the north monopole is compressed to form the flux tube, through which the south monopoles are confined via color-like force[2,3-6] and generated the torus sublattice. For example, two monopoles, **S5** and **S7** are connected through the flux tube (**N6**) and three monopoles are connected through the two flux tubes as shown in Fig. 2 (b) and (c), respectively. Thus contrary to Dirac's monopole as drawn in Fig. 2(a), in this case there is no Dirac's string singularity, indicating the zero flux leakage. Which is consistent with 't Hooft's original idea of monopoles with compact space(s) and formation of the torus sublattice[2]. This is important to note that in Fig. 2(b) and (c) monopole **S5** and **S3** has higher power in terms of broadening (not eigenstate) compared with their parent counterparts, which is not happened in case of Dirac's monopoles (Fig. 2(a)) (see text for detail). Which is due to accumulation of energy of defect north monopole(s).

Our experimental setup consisted of a simple configuration as shown in Fig. 3. Initial system of lattice solitons are named as monopoles, as depicted schematically in Fig. 1., and experimentally in Fig. 4(a, c). The input system consists of two lattices whose wavelength range is from 999.5 to 1009.5 nm, however the unaffected mode at ~1006 nm in between the two lattices representing a kind of isolation between the two lattices (for detail see the figure caption). This also shows that presence of more than one modes in one wave packet is required to achieve graphene-like resonance. The initial lattices are presented in Fig. 4(a) and (c). At the first (50:50) coupler the input lattices power splitted into two copies. These copies meet resonantly at the second (50:50) coupler (see methods section) and ref [20]. The resonance generated pseudomagnetic flux between the monopoles owing to radiation loss at the (so-called) junctions. The direction of the mild fluxes or radiation are represented by drawing thin blue lines in the Fig. 4(b). Experimental results are captured using optical spectrum analyser (OSA) and presented in Fig. 4(b) and (d). In Fig. 4(b), Dirac's monopoles are **S1 and N1**. Usually, it is believed that Dirac's string could not be detected in any experiment[1,24], however, in our experiment monopoles confined electromagnetic field i.e., photons. Therefore, probably our observation is due to escaping (leaking) of the electromagnetic field along with magnetic flux leakage. This description may also be understandable using description of event horizons at the interfaces of hyperbolic metamaterials[25-26]. Monopoles, **S5, N6, S7** are transformed into 't Hooft's subgroup or sublattice, where **N6** monopole is compressed and form the flux tube, through which **S5** and **S7** are connected via color-like force. The bottom of **S5** is broadened due to accumulation of **N6** power. The direction of the transferred power is an indicator of the flux tube direction or direction of color-like force, as shown in experimental results depicted in Fig. 4 (b). Another 't Hooft's sublattice consisted of four monopoles (**S2**, **N3, S4**, **N4**), where a defect monopole, **N3** transformed into a flux tube. Surprisingly in this case a south, north monopole-symmetric pair (**S4, N4**) is also a part of the sublattice, however, seemingly they jointly formed a 't Hooft's monopole as evident with the outer tails of **S4** and **N4** in Fig. 4(b). We observed that unlike Dirac's north monopole (**S1)**, the 't Hooft's monopole did not allow leaking of magnetic flux (Dirac string), therefore no electromagnetic flux leakage is observed. The space with leakage and the compactness of space has close relevance with quantization of magnetic and electrical charges as $ge = 1/2$ and $ge = 1$ in Dirac's and 't Hooft's space, respectively[1-2]. In case of second input lattice (Fig. 4(b)), a 't Hooft's sublattice with two flux tubes was observed. Here, the 't Hooft's sublattice consisted of two defect north monopoles (**N2**, **N4**), one defect south monopole (**S3**) and two boundary monopoles (**S1**, **S5**). The defect monopoles (**N2**, **N4**) generated two flux tubes. However, for convenience of understanding this sublattice could be considered as a combination of further two sublattices as indicated with two dotted rectangular boxes in Fig. 4(d). Straight line extension of the tube is clearly observed at the location of **N6, N3** in Fig. 4(b) and at **N2** in 4(d). These experimental observations are agreed with t' Hooft's[2] and others theories[3-4,23] of solitonic monopoles. Thick red and blue arrows near the wavelength axis are drawn to show red and blue shifting of the lattices with respect to parent lattices. It seems that the shifting is due to straightening and direction of the flux tubes, from **S2 to S4 and S7** to **S5** in Fig. 4(b) and **S1** to **S3** in Fig. 4(d). However, strength of the eigenstate, 0.038 of 't Hooft's monopole (**S3**) is preserved, which is equivalence of corresponding input monopole (**S3**). In one sense this indicates complete preservation of the input eigenstate only when both sides of **S3** is surrounded by the flux tubes[27]. However, indeed, the bottom of the 't Hooft's monopole (**S3**) is broadened however, no effect on triangular shape of the parent **S3** monopole and 't Hooft's monopole. Which indicates accumulation of the radiation power of **N2** and **N4** monopoles. However, no change in the shape indicates that there is nothing like conventional photonic transition (evanescent) instead monopoles radiate to generate magnetic field and form continuum[20].

Summary and conclusions: Indeed intrinsically, Berry's, EP-induced and topological-phase transitions are related with Dirac's and 't Hooft's monopoles[2,28-29]. In this manuscript, we provided some of the experimental evidencs. From resonances (anti-crossings) of initial monopoles (Dirac's cones) and generation of the radiation from Dirac points[20], we experimentally observed transformation of monopoles into 't Hooft's torus sublattices, Dirac's north and south monopoles. For Dirac's monopoles, the space has the flux leakage. However, defect monopoles completely closed the flux leakage (compact space), and the 't Hooft's monopoles and the torus sublattices are observed. The preservation of shapes of parent and final monopoles indicated the relevance with Bogomolny vortices[18-19] ($\left(k=k_c=1/\sqrt{2}\right)$). Bogomolny vortices are dual of Abrikosov vortices[30-31] ($\left(k>1/\sqrt{2}\right)$). Bogomolny vortices and their transformation into monopoles and chromoelectric flux tubes are reviewed in the context of magnetic monopoles and solitons boson-fermion duality or supersymmetry[3-6,32]. It seems that in our experiments, the compact and leakage spaces were created in Hilbert spaces, which might helpful to initiate an experimental field where the flux can evolve lower dimensions into higher dimensions. We believe, our results are importance for further understanding and experiments of decoherence free quantum computation, color confinement or dual superconductivity, cosmic strings, gravity and condensed matter-like multimode interference.

**METHODS SUMMARY**

The two soliton lattices occupying wavelength range from 999.5 to 1,009.5 nm (whose schematic presentation is shown in Fig. 1.) is entered in two metre loop of optical fiber through the fibre dense wavelength division multiplexer (DWDM). The entire loop consists of a normal dispersion optical fibre (HI 980). The fibre exhibits zero dispersion at 1,300 nm, and the dispersion values at 980 nm are approximately -63 ps/nm/km. The fibre exhibited multimode character. Normal dispersion fiber is used to avoid modulation instability. Two copies of the input lattices with attenuated power are generated after meeting with the first 50:50 coupler. The optical spectrum of the lattices detected at the output of second coupler is captured using optical spectrum analyser (OSA). First the lattices are detected without any change in the structure due to absence of interference/resonance at the second coupler and depicted in Fig. 4(a) and (c). Detected peak powers of the lattices are attenuated to 3.38 and 1.15 μW as shown in Fig. 4(a) and (c), respectively. This detection is important for comparing the final resulted, consisted of Dirac's monopoles and 't Hooft's monopoles. Graphene-like resonance is achieved by resonances of power splitted lattices at the second coupler by exactly matching the length of the two arms of the loop as shown in Fig. 3. The resonances generated Dirac's monopoles and 't Hooft's torus sublattices and the results are shown in Fig. 4 (b) and (d). For further details see our recent paper, ref [20].

**Acknowledgements**

This work was partially supported by the "Systems biology infrastructure establishment grant" and the "Ultrashort Quantum Beam Facility Program" through a grant provided by the Gwangju Institute of Science and Technology in 2015.

**Author Contributions**

M.I. conceived the idea, designed and performed the experiments and the analysis of results. Y.T. provided insightful advice and encouraged M.I. for the investigation and supervised the project. M.I., H.S. and Y.T. wrote the manuscript. All authors reviewed the manuscript.

**Competing Interests Statement**

The authors declare that they have no competing financial interests.

Figure captions

**Figure 1:** Schematic of input system of monopole lattices. **a, b,** The modulated solitons have uniform spatial gap equal to 0.54 nm. Phase modulation between any two solitons is calculated with reference to a tilt (dotted line). **a,** In total the solitons in the lattice are representing nine monopoles. Where the monopoles following symmetry of north (**N**), south (**S**) monopoles are represented with green lines. The boundary modes are represented with orange lines. Defect north monopoles are represented with red lines. **b,** In total this lattice is consisted offive monopoles. Where two defect north monopoles, **N2** and **N4**, one defect south monopole, **S3 and** two boundary monopoles, **S1** and **S5.** Defect south monopole is represented with purple line. Inset indicates coordinates of the lattices.

**Figure 2:** Schematics of selected results generated from the input system, containing depiction of Dirac's monopoles and Dirac's string, and 't Hooft's monopoles, sublattices (subgroups) and the flux tubes. **a,** The black curve symbolizes a Dirac's string due to magnetic flux leakage from the monopole. **b,** 't Hooft's torus lattice having two monopoles and one chromoelectric-like flux tube. There is no string singularities because 't Hooft's monopoles are of compact space therefore do not allow escaping of magnetic flux. Moreover, energy of **N6** defect monopole is transferred to **S5** monopole without effecting the relative eigenstate of **S5** and form the flux tube. Red arrows in (**b** and **c**) indicate the directions of energy transfer as well as direction of the flux tube(s). **c,** Three monopoles (**S1**, **S3**, **S5**) connected through the two flux tubes. Where energy of defect north monopoles, **N2** and **N4** are transferred to defect

south monopole **S3** without effecting original eigenstate of **S3** monopole. The **N2** and **N4** are transformed into the two flux tubes. Even though eigenstates of **S3 (b)** and **S4 (c)** are not affected, however energy increments are evident with broadening of the corresponding 't Hooft's monopoles.

**Figure 3:** Experimental setup. The two lattices ranging from 999.5 to 1,009.5 nm (whose schematic presentation is shown in Fig. 1.) is entered in two metre loop of optical fiber through the fibre dense wavelength division multiplexer (DWDM). The entire loop consists of a normal dispersion optical fibre (HI 980). The fibre exhibits zero dispersion at 1,300 nm, and the dispersion values at 980 nm are approximately -63 ps/nm/km. The input lattices are power splitted into two copies after meeting the first 50:50 coupler. The optical spectrum of the lattices detected at the output of second coupler is captured using optical spectrum analyser (OSA). First the lattices are detected at the output of the setup without any change in the structure, however, the structure is attenuated (power splitted), as shown in Fig. 4(a) and (c). Anti-crossing of monopoles are achieved by graphene-like resonance at the second coupler. Consequently, Dirac's monopoles and 't Hooft's torus sublattices are generated. The inset shows symbolic representation of graphene-like resonance at coupler.

**Figure 4:** Experimental results. Monopole lattices detected at the output of second coupler of the experimental setup, before and after generation of Dirac's monopoles and string singularity, and 't Hooft's monopoles and torus sublattices. **a, c,** First and second lattice are consisted of nine and five monopoles, respectively (for detail see Fig. 1). **b, d,** Thin blue arrows are drawn to show the direction of radiation loss and direction of flux owing to phase contrast. Dirac's north and south monopoles, string are generated from the monopoles having north, south symmetry (**S1**, **N1**). Blue arrow near **S1** is drawn to show the leakage of flux or radiation. 't Hooft's monopoles, the flux tubes and torus sublattice are generated owing to defect monopole(s). Red arrows indicate directions of the flux tubes as well as power transfer from defect **N** monopole to **S** monopole in the respective sublattices. The gap between **S1** and **S3** in (d) is become greater than parent gap (2*0.54 nm) of corresponding monopoles due to straight

extension of the tube. Thick red and blue arrows is drawn to show red and blue shifting of the lattices with respect to the parent lattices. The torus sublattices are encircled with dashed lines rectangular boxes. This is extremely important to note that in case of Dirac's north monopole (**S1**), electromagnetic field leakage is observed as string singularity owing to magnetic flux leakage as evident with an oblong-shape at **S1**. Near **S1**, a blue arrow indicates direction of flux leakage. Contrarily, in case of 't Hooft's monopoles, **S7** in (b), and **S1** and **S5** in (d), no (oblong-shape) leakage of the flux or electromagnetic field is observed. These observations are consistent with Dirac's and 't Hooft's original ideas. **b, d,** The broadening of the boundary monopoles (**S7**, **S1**, **S5**) is due to their presence at the end of the lattices. **d,** Here, a big torus lattice is observed having three monopoles and the two flux tubes. This 't Hooft's torus sublattice can be visualized as two sublattices, which are shown and highlighted in the figure by encircling with the two rectangular boxes. (For further detail see caption of Fig. 2 and the article text) A mode represented as 'unaffected mode' located at 1006 nm wavelength in **c**, and **d**, is unaffect by the transformations. Which indicates the isolation of first and second lattice from each other.

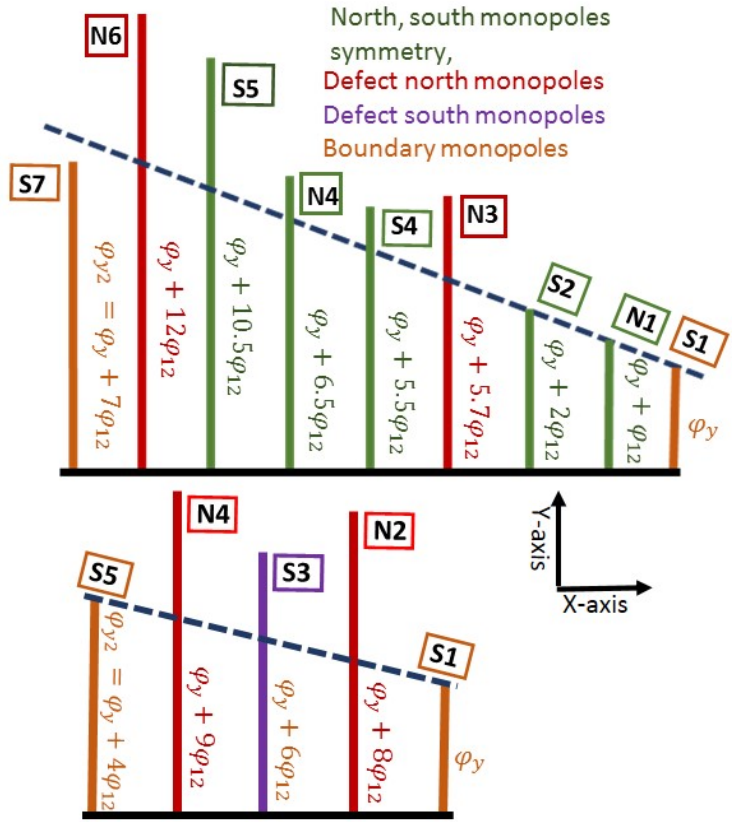

**Figure 1**

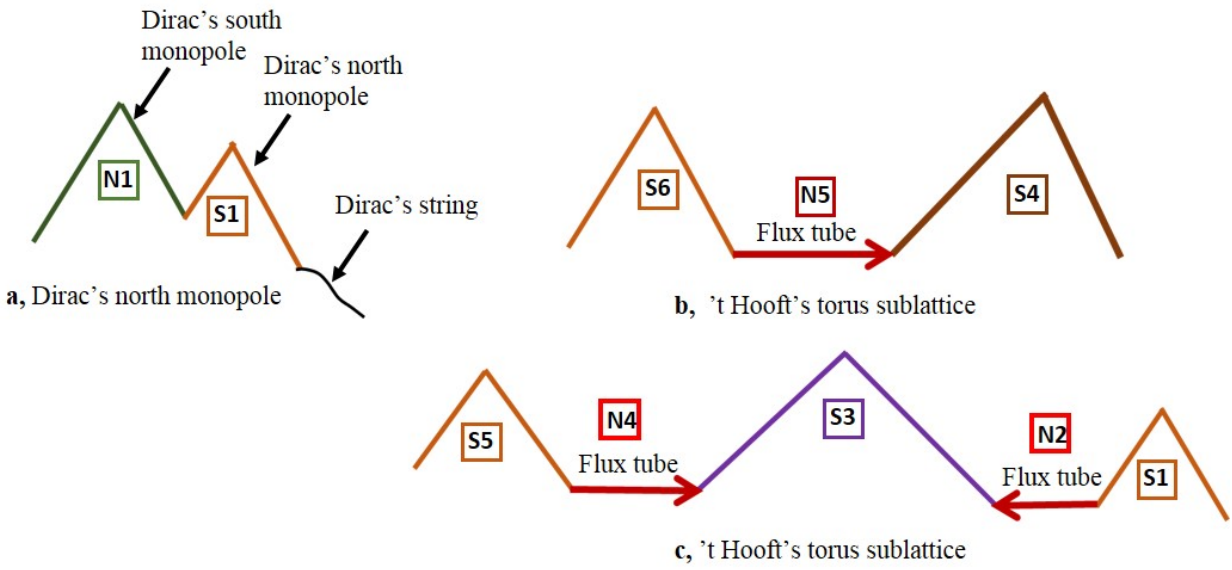

**Figure 2

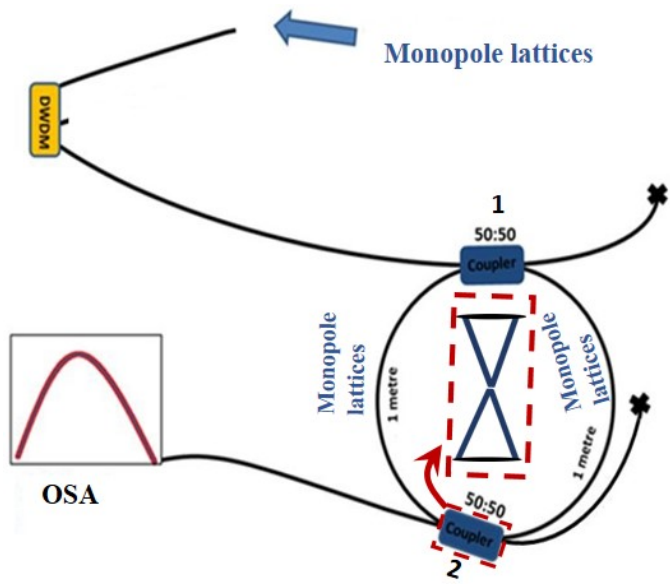

**Figure 3**

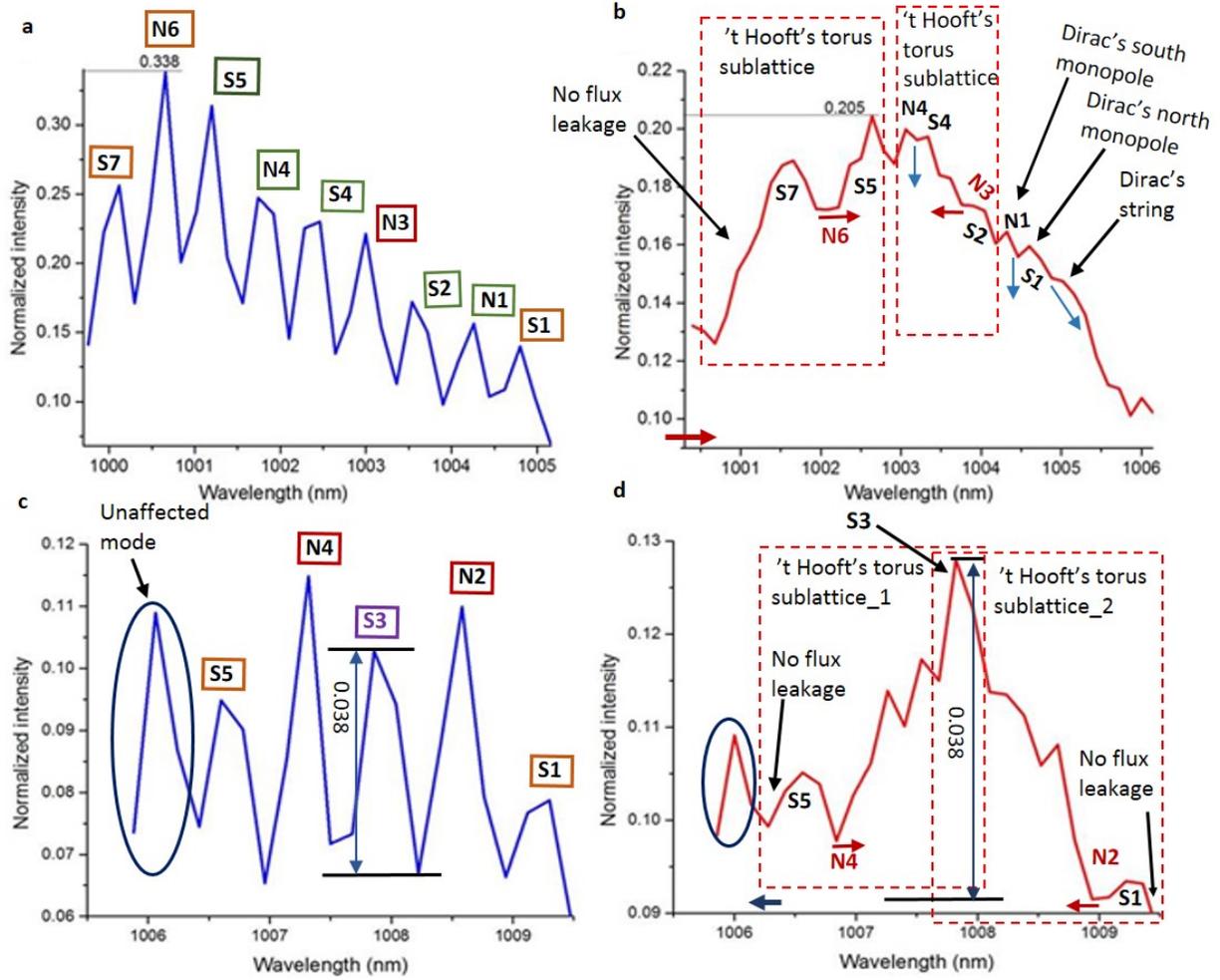

**Figure 4**